**Prospects of Single-Cell NMR Spectroscopy with Quantum Sensors**

Authors: Nick Ruben Neuling[a], Robin Derek Allert[a], Dominik Benjamin Bucher[a,b]

[a]Technical University of Munich, TUM School of Natural Sciences, Department of Chemistry, Lichtenbergstr. 4, 85748 Garching b. München, Germany

[b]Munich Center of Quantum Science and Technology (MCQST), Schellingstr. 4, 80779 München, Germany

Correspondence concerning this article should be addressed to D.B. Bucher at dominik.bucher@tum.de

**Abstract**

Single-cell analysis can unravel functional heterogeneity within cell populations otherwise obscured by ensemble measurements. However, non-invasive techniques that probe chemical entities and their dynamics are still lacking. This challenge could be overcome by novel sensors based on nitrogen-vacancy (NV) centers in diamond, which enable nuclear magnetic resonance (NMR) spectroscopy on unprecedented sample volumes. In this perspective, we briefly introduce NV-based quantum sensing and review the progress made in microscale NV-NMR spectroscopy. Lastly, we discuss approaches to enhance the sensitivity of NV ensemble magnetometers to detect biologically relevant concentrations and provide a roadmap towards their application in single-cell analysis.

**Keywords**

Nuclear Magnetic Resonance, Nitrogen-Vacancy Center, Single-Cell Analysis, Metabolomics, Hyperpolarization, Quantum Sensing, Lab-on-a-Chip

**Introduction**

Phenotypic variations of individual cells are present in all biological systems, even in clonal or isogenic cultures [1]. These cell-to-cell heterogeneities have high biological relevance, for instance, in developing antibiotic resistance [2], carcinogenesis [3], and even immune responses [4]. Most single-cell studies investigate genetic information using advanced single-cell sequencing techniques (genomics and transcriptomics), which rely on amplifying nucleic acids to increase signals for detection [5]. This method of signal amplification is impossible for proteins and metabolites, making single-cell proteomics and metabolomics more challenging since the absolute molecular numbers in single cells can be very low [6]. In contrast to proteins, small molecules such as metabolites usually cannot be labeled with dyes and do not exhibit autofluorescence, which often hampers optical detection approaches. Moreover, the metabolome reacts in a very sensitive and fast way to environmental influences, which not only makes its analysis exceedingly difficult, but it also highlights the importance of metabolomics to understand a cell's phenotype [1].

Despite these challenges, some methods, such as mass spectrometry (MS) and Raman microscopy, can detect metabolites in single cells [7–9]. While MS is powerful in sensitivity and specificity, its invasiveness precludes applications such as long-term in-vivo studies. Raman microscopy, although non-invasive, has limited molecular specificity due to broad and overlapping spectral lines.

Nuclear magnetic resonance (NMR) spectroscopy is non-invasive and arguably the most powerful method for chemical analysis. However, due to its low intrinsic sensitivity, conventional NMR spectroscopy requires macroscopic sample volumes of up to several hundred microliters, impeding its use for single-cell analysis (see Figure 1a) [10]. This limitation could be overcome by using microcoils as NMR detectors, which have been shown to achieve the sensitivity for detecting NMR signals from cells down to sub-nanoliter volumes (e.g. ova of the tardigrade *Richtersius coronifer* and early cow embryos) [11–13]. Although impressive, these experiments are limited by i) the detection length scale which is still around 100 μm and ii) the lack of spatial resolution or imaging capabilities. These limitations can be overcome by a novel NMR sensor based on quantum defects in the diamond lattice [14,15]. This approach has demonstrated NMR spectroscopy at unprecedented length scales, ranging from picoliter volumes [16–19] down to a single molecule [20] and even down to a single spin [21,22], enabled by their atomic size and optical readout.

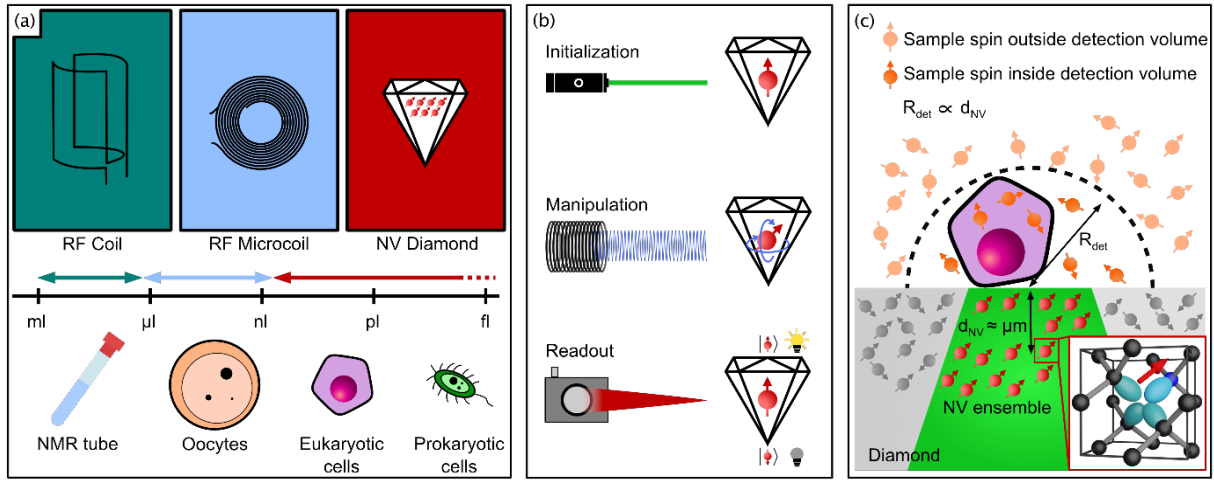

**Figure 1. Basic principles of microscale NMR spectroscopy with NV centers in diamond.** a) Samples and sample volumes of different NMR techniques. Conventional NMR requires several hundred microliters of sample volume, whereas NMR of nanoliter volumes (e.g., single oocytes) has been demonstrated using microcoils [11–13]. Nitrogen-vacancy (NV) centers in diamond enable NMR spectroscopy of picoliter sample volumes (corresponding to small eukaryotic cells) [16–19]. b) Illustration of the NV center properties enabling quantum sensing. The quantum state (i.e., spin state) of the NV center can be optically polarized, manipulated with microwave (MW) fields, and optically read out by its spin state dependent fluorescence. c) For NV ensembles the radius $R_{det}$ of the detection volume (dotted lines) corresponds to the average NV center depth $d_{NV}$. The inset shows the NV center structure in the diamond lattice with the vacancy in the center and the nitrogen atom in blue.

**NMR spectroscopy with nitrogen-vacancy centers in diamond**

The nitrogen-vacancy (NV) center is a spin defect in the diamond lattice which is formed by a nitrogen (N) impurity next to a missing carbon (vacancy, V) and has attracted intense interest as a quantum sensor in recent years [23,24]. Although these defects occur naturally, NV centers for sensing applications are usually artificially manufactured, either by nitrogen ion implantation or by adding nitrogen during diamond growth by chemical vapor deposition [25]. For sensing applications, the negatively charged $NV^-$ is used (for the following denoted as NV), where six electrons occupy four orbitals around the vacancy, resembling electrons bound to an atom's nucleus. This system has well-defined optical transitions and long-lived electronic spin states at room temperature with three essential properties which form the basis for its quantum sensing capabilities (Figure 1b): i) the NV-spin state can be initialized optically by laser excitation, ii) the NV-spin state can be coherently manipulated with resonant microwave (MW) fields, and iii) the NV-spin state can be read out optically by its spin state dependent fluorescence, which enables optically detected magnetic resonance (ODMR) experiments [26]. This atom-like system enables the realization of a fully functional qubit, which interacts with its environment (i.e., the magnetic field of an NMR signal) and results in an optically readable signal (for details see Box 1). An advantage of the NV center is that it can be brought close (nano- to micrometer) to the diamond surface and thus, the NMR sample. As a rule of thumb, the radius of the detection volume corresponds roughly to the distance of the NV center to the diamond surface (single NV center) or the average NV center depth (NV ensemble) and, accordingly, enables NMR spectroscopy at the nano- to microscale (Figure 1c) [27–31]. For more details, interested readers are referred to a recent review on nano- and microscale NV-NMR by Allert *et al.* [28].

The NV center has an electronic spin triplet ground state with the $|m_s = 0\rangle$ ($|0\rangle$) state and the degenerate $|m_s = \pm 1\rangle$ ($|\pm 1\rangle$) states separated by a zero-field splitting of 2.87 GHz. For sensing applications, an external magnetic field is applied along the NV symmetry axis which lifts the degeneracy of the $|\pm 1\rangle$ states. By addressing the $|0\rangle \rightarrow |1\rangle$ or the $|0\rangle \rightarrow |-1\rangle$ transition, a fully functional qubit is realized, which can be represented on a Bloch sphere where the poles correspond to the basis states, e.g., $|0\rangle$ and $|1\rangle$. The Bloch vector (red arrow) indicates the qubit's spin state. In a typical quantum sensing experiment, the NV spin state is initialized to the $|0\rangle$ state by a laser excitation pulse (green rectangle) and subsequently transformed into the active sensing state, e.g., a superposition state $|\Psi\rangle = 1/\sqrt{2}\,(|0\rangle + |1\rangle)$, by a MW $\pi/2$ pulse (blue rectangle). Then, the superposition state is allowed to evolve under the influence of the magnetic field to be sensed ($|\Psi\rangle = 1/\sqrt{2}\,(|0\rangle + e^{-i\Phi}|1\rangle)$), which leads to an accumulation of a relative phase $\Phi = \int_0^t \gamma_{NV} B(t) dt$, where $\gamma_{NV}$ is the gyromagnetic ratio and B(t) the signal magnetic field. Multipulse (dynamic decoupling) sequences are used to detect oscillating signals (ac signals), such as NMR signals created by precessing nuclear spins (orange sinusoid). These sequences consist of evenly spaced MW $\pi$ pulses, where the phase accumulation is maximized when the MW pulses coincide with the nodes of the ac signal [32]. By sweeping the pulse spacing, a spectrum of the ac signal can be recorded where the spectral resolution is limited by the coherence time of the NV center. After evolution, the coherence of the NV spin state is transformed into a measurable population by another MW $\pi/2$ pulse. Finally, the NV spin state is read out by its spin state dependent fluorescence with probability $P_{|0\rangle} = |\langle 0|\Psi\rangle|^2 = \sin^2(\phi/2)$. This scheme forms the basis for most NMR sensing approaches. For microscale NV-NMR the coherently averaged synchronized readout (CASR) pulse sequence is typically used which overcomes the spectral resolution limit imposed on commonly used NV sensing schemes caused by NV decoherence and short spin state lifetimes [16]. In the CASR scheme, the multipulse sequences are synchronized with the external NMR signal which results in an oscillation of the NV fluorescence over subsequent optical readouts (red sinusoid) and resembles the free induction decay (FID) in conventional NMR spectroscopy. We would like to direct interested readers for details about quantum sensing to the excellent review by Degen *et al.* [32].

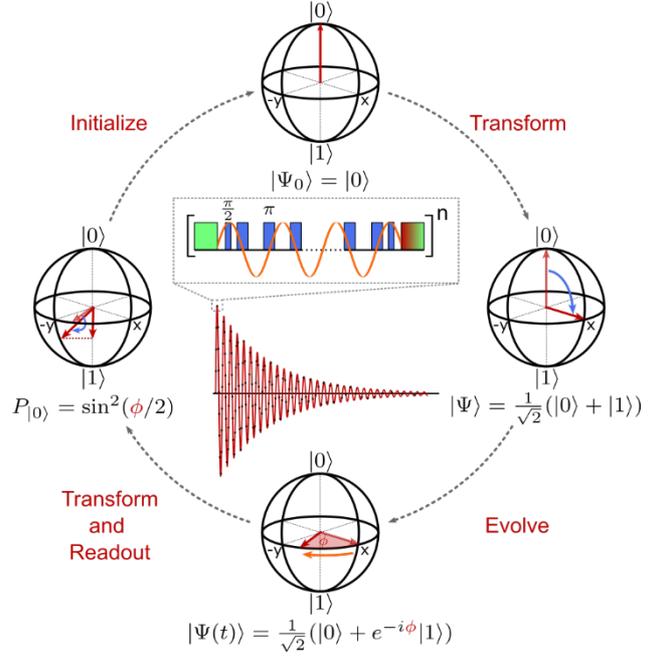

**Box 1. Quantum sensing with NV centers in diamond.**

## State-of-the-art microscale NV-NMR

In 2018, Glenn *et al.* demonstrated microscale NV-NMR for the first time [16]. Whereas most of the previous NV-NMR experiments had been performed with single NV centers for nanoscale detection [27,30], the authors showed that an ensemble of NV centers in a micrometer-thick layer can be used to detect NMR signals from picoliter sample volumes (~$(10\ \mu m)^3$ comparable to the volume of a single eukaryotic cell). Crucially, they developed a pulse sequence called coherently averaged synchronized readout (CASR) that can provide high-resolution NMR spectra with resolved scalar couplings and chemical shifts, as demonstrated for several organic compounds. Furthermore, they achieved a proton number limit of detection (nLOD) of $3\ nmol/\sqrt{Hz}$ (i.e., $300\ M/\sqrt{Hz}$ concentration limit of detection, cLOD), which is defined as the minimum number of sample nuclear spins needed to obtain a signal to noise ratio (SNR) of 3 in 1 s of averaging time [33]. Although these experiments enabled probing of picoliter volumes, the measured concentration sensitivity was still poor and limited to samples with high molarity. One reason was the low applied magnetic field (88 mT) which led to poor thermal polarization of the proton nuclear spins (~$3\times10^{-5}$ % according to the Boltzmann distribution), thus limiting the measurable NMR signal. Smits *et al.* [19] combined microscale NV-NMR with direct flow-based prepolarization at 1.5 T, achieving an nLOD of $1.1\ nmol/\sqrt{Hz}$ within a sensing volume of 40 picoliters (i.e., $^1$H cLOD of $27\ M/\sqrt{Hz}$). In addition, the authors demonstrated two-dimensional NMR spectroscopy, an important technique used in conventional NMR to assign resonances in complex molecular structures or mixtures unambiguously.

Further signal enhancements are possible using hyperpolarization techniques, which can boost nuclear spin polarization up to several orders of magnitude beyond Boltzmann polarization. Bucher *et al.* [17] applied Overhauser dynamic nuclear polarization (ODNP), which takes advantage of the higher electronic spin polarization of stable radicals added to the sample to increase the proton spin polarization by up to two orders of magnitude. With this, the authors achieved ~6×10$^{-3}$% nuclear spin polarization at 85 mT with an nLOD of $10\ pmol/\sqrt{Hz}$ ($1\ M/\sqrt{Hz}$). However, the achievable nuclear spin polarization is limited by the Boltzmann polarization of the electronic spins at ambient temperatures and low magnetic fields (below 1 T) typically used in NV-NMR. Thus, Arunkumar *et al.* [18] combined NV-NMR with a parahydrogen induced polarization (PHIP) technique termed signal amplification by reversible exchange (SABRE). This technique transfers the singlet spin order of parahydrogen to the sample nuclear spins which can produce a polarization of over 10% [34]. In their study, the authors achieved nuclear spin polarization levels of 0.5%, allowing them to detect pyridine and nicotinamide in low millimolar concentrations within a ten picoliter detection volume, achieving a record nLOD of $66\ fmol/\sqrt{Hz}$ ($6.6\ mM/\sqrt{Hz}$).

However, to study single cells and their response to external stimuli, both high sensitivities and the ability to manipulate cells in time and space are required. Fortunately, this problem can be solved by microfluidics, for which various control and cell-trapping methods have been developed in the last decades [35]. Allert *et al.* have shown that NV-based quantum sensing can be interfaced with microfluidics in a biocompatible manner, allowing for simultaneous positioning and imaging of NMR samples or cells on the NV-diamond surface, an essential step towards single-cell studies [36].
We would like to note that NMR spectroscopy covers a wide range of techniques with diverse applications. Recently, the capabilities of NV-NMR experiments have been extended by the use of pulsed field gradient techniques, which allow to study water diffusion within individual microstructures [37]. These results lay the groundwork for future studies on the diffusion of water or metabolites in single cells.

**Technological advances and estimated sensitivity improvements**

In the following, a possible roadmap for single-cell NV-NMR is presented by discussing technological advances in the field of microscale NV-NMR and the associated improvements in sensitivity using the work of Glenn *et al.* [16] (Boltzmann polarization) and Arunkumar *et al.* [18] (hyperpolarization) as starting points. If not specified, nLOD and cLOD are referred to as LOD. The results are depicted in Figure 2 and summarized in Table 1. Figure 2 correlates the achieved and projected nLOD with the detection volume. In addition, it displays the cLOD as grey diagonal lines. The colored dashed diagonal lines indicate how the LOD would change with detection volume, corresponding to a change in sensor volume $V$, i.e., the overlap of the excitation laser and the NV layer thickness and, therefore, the number of probed NV centers. As Bruckmaier *et al.* [31] discussed in detail, the NMR signal strength does not change if the sensor-to-sample geometry remains constant. However, due to the scaling of the number of interrogated NV centers, the sensor's sensitivity changes with $\sqrt{V}$. The nuclear spin number is linearly dependent on the detection volume at a fixed concentration, hence $nLOD \propto \sqrt{V}$ and $cLOD \propto 1/\sqrt{V}$. Finally, in analogy to Eills *et al.* [38], the green patch with a boundary at $cLOD = 5\ mM/\sqrt{Hz}$ indicates where single-cell metabolomics is proposed to be feasible. An extensive overview of cellular metabolite concentrations in mammalian iBMK, yeast, and E. coli cells is provided by Park *et al.* [39]. Metabolites of particular interest, such as amino acids, pyruvate, and fumarate, have cellular concentrations in the range of 10$^{-4}$ to 10$^{-2}$ M.

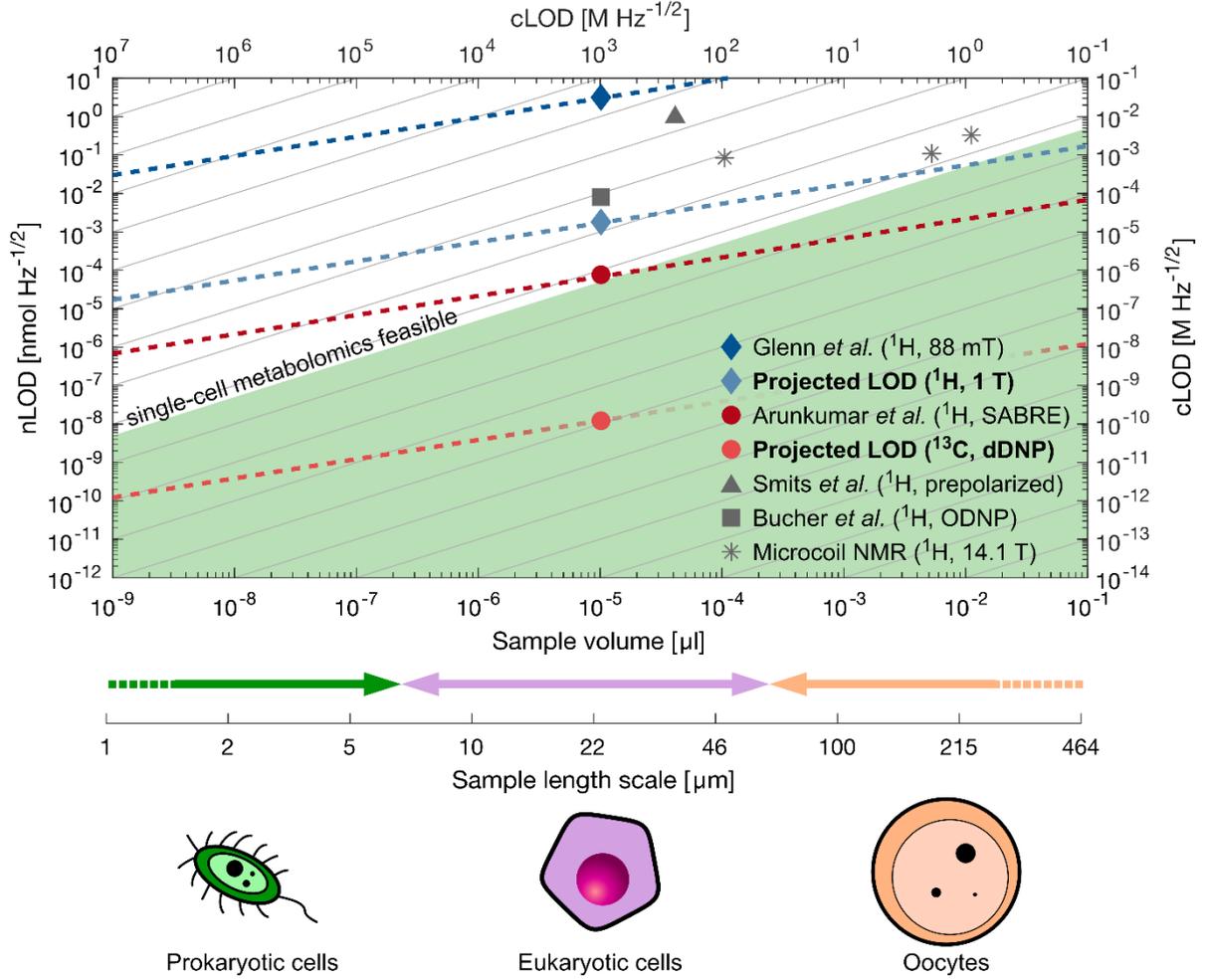

**Figure 2. Achieved and projected LOD correlated with the detection/sample volume and sample length scale.** The cLOD is displayed as grey diagonal lines. The colored dashed diagonal lines indicate how the LOD would change with detection volume, corresponding to a change in sensor volume $V$, i.e., the overlap of excitation laser and the NV layer thickness and, therefore, the number of NV centers. The green patch indicates where single-cell analysis is proposed to be feasible [38]. Projected LOD (light blue rhombus, $^1$H, 1 T) for untargeted analysis takes into account improved NV readout fidelity (~30 fold), improved diamond material (~5 fold) and increased magnetic field (~11 fold) in comparison to values reported in Glenn *et al*. (dark blue rhombus, glycerol, $^1$H, 88 mT) [16]. Projected LOD (light red circle, $^{13}$C, dDNP) for the targeted approach accounts for improved NV readout fidelity (~30 fold), improved diamond material (~5 fold) and hyperpolarization techniques (~35 fold detecting $^{13}$C, lower gyromagnetic ratio considered for calculation) in comparison to values reported in Arunkumar *et al*. [18] (dark red circle, pyridine, $^1$H, 6.6 mT, SABRE). Grey markers represent the works of Smits *et al*. (triangle, water, $^1$H, 13 mT, prepolarization at 1.5 T) [19], Bucher *et al*. (square, water, $^1$H, 85 mT, ODNP) [17] and microcoil NMR (asterisks, $^1$H, normalized to 14.1 T, demonstrations of magnetic resonance imaging not included) [33].

Gains in LOD can be achieved either by improving the NMR signal characteristics or the NV sensor's (ac) sensitivity. The latter one can be expressed for an ensemble of NV centers at the shot-noise limit [40]

$$\eta_{shot} \propto \frac{1}{C\sqrt{N_{NV} n_{avg} T_2}}, \qquad (1)$$

where $C$ is the spin state dependent fluorescence contrast, $N_{NV}$ is the number of NV centers, $n_{avg}$ is the average number of photons collected per NV, and $T_2$ is the NV's coherence time. The sensitivity is defined as the minimum detectable magnetic field per second averaging time; therefore, reducing $\eta_{shot}$ increases the magnetic field sensitivity. Equation 1 indicates different ways to improve sensitivity. First, the readout fidelity of the NV spin state can be improved by increasing the photon collection efficiency (see Table 1). We anticipate a larger improvement by implementing quantum logic enhanced (QLE) readout schemes. The information on the NV spin state is lost during the readout process due to concomitant repolarization into the $|0\rangle$ state. QLE protocols use long-lived nuclear ancillary spins as memory qubits that store the NV electronic spin state for subsequent repetitive readouts, improving the readout fidelity. Although not yet demonstrated with the CASR pulse sequence, we expect that QLE readout schemes will improve sensitivity by an order of magnitude [41]. Secondly, the contrast C, coherence time $T_2$ or $N_{NV}$ could be increased by improving the diamond matierial. Generally, the creation efficiency of negatively charged NV centers in diamond is low (typically ~ 5%) [42,43]. One promising solution is the additional incorporation (co-doping) of other non-nitrogen-based n-type dopants, such as sulfur or phosphorous [44–46]. Creation efficiencies > 50% could be achieved [47], which would result in a factor of ~ 3 in sensitivity improvement. In addition, phosphorous co-doping can increase the coherence properties of NV centers by a factor of ~ 2 [45], corresponding to a factor of ~1.4 in sensitivity improvement. Therefore, diamond material engineering could yield an overall improvement of ~ 5 for NV-NMR spectroscopy. The research area of NV diamond material science is still full of untapped potential, and we expect major advancements in the mid to long term.

On the NMR side, we envision substantial improvements in LOD. Microscale NV-NMR would benefit greatly from higher magnetic fields, as they increase the Boltzmann nuclear polarization and most importantly, the spectral resolution. However, NV sensing of NMR signals at high magnetic fields (>> 1 T) is challenging due to the increased complexity of the required high-frequency MW equipment and technology. So far, all microscale NV-NMR experiments have been performed at low magnetic fields (< 0.2 T), but we expect that with current sensing schemes and appropriate hardware, magnetic fields of 1 T can be reached. Alternative sensing schemes have been proposed to work at even higher magnetic fields but have yet to be demonstrated experimentally [48]. Glenn *et al.* [16] obtained relatively broad resonance lines of approximately 9 Hz, which limits chemical specificity and the LOD. Generally, reducing the NMR linewidth increases the SNR and consequently enhances the LOD. As the $T_1$ relaxation time dictates the repetition rate for signal averaging in experiments, the LOD is sample-dependent (e.g. Glenn *et al.* [16] determined their nLOD on glycerol), and a generalization is omitted in this context. Nonetheless, narrowing the NMR linewidth is a crucial objective in microscale NV-NMR. Thus far, a thorough investigation of the NV-NMR linewidth limit is still lacking and needs further study. However, the implementation of a highly homogeneous magnetic field and shimming techniques [49] should reach benchtop NMR standards, featuring a linewidth of less than 0.5 Hz at 1 T. These improvements will certainly enable the study of water diffusion at the a single cell level. However, even with these improvements, the predicted LOD would still be nearly two orders of magnitude above the stated threshold for detecting metabolites in single mammalian cells (Figure 2). While signal averaging allows to reach this threshold, the ability to resolve dynamics or metabolic fluxes may be lost. For that reason, the detection of a single-cell's metabolic fingerprint, referred to as "untargeted approach", is limited to the detection of highly concentrated metabolites with reduced temporal resolution. Nevertheless, this fingerprinting may have important applications, particularly in combination with machine learning approaches to classify cell phenotypes [50].

Another way to increase the nuclear spin polarization and, therefore, the LOD by several orders of magnitude is to employ hyperpolarization techniques, as discussed before. Here, isotopically labeled metabolites are typically hyperpolarized and fed to the cell, which we refer to as "targeted approach". The advantages are the background-free detection of hyperpolarized analytes and the ability to monitor metabolic fluxes. Moreover, there is no need for high magnetic fields, which reduces technical complexity and minimizes magnetic susceptibility mismatches within biological environments. We think that PHIP [51] and dissolution dynamic nuclear polarization (dDNP) [52] are the most promising polarization schemes for NV-NMR. Especially the field of PHIP made rapid progress in recent years, providing chemically pure solutions of metabolites with polarization levels of up to 25% $^{13}$C polarization after purification, competitive with dDNP [53]. Polarization transfer in dDNP experiments occurs at cryogenic temperatures and high magnetic fields between MW-irradiated radicals added to the sample and the sample nuclear spins. After polarization, the sample is rapidly melted (dissolution) and transferred to the measurement apparatus [54]. So far, dDNP offers the highest $^{13}$C polarization levels with up to 70% $^{13}$C polarization after dissolution but has yet to be interfaced with NV-NMR [55]. By implementing state-of-the-art technological advances, hyperpolarized NV-NMR could reach attomol nLOD at picoliter volumes corresponding to micromolar cLOD (Figure 2). We envision that hyperpolarization enhanced NV-NMR paired with microfluidics will enable single-cell analysis of small eukaryotic or even prokaryotic cells with the benefit of detecting metabolism in real-time.

**Table 1. Overview of approaches to improve LOD.**

| | Parameter | Method | Method description and evaluation | LOD improvement in comparison [16] and [18] | Validity for single-cell studies |
|---|---|---|---|---|---|
| NV sensor | NV readout fidelity | Improved fluorescence collection | Improves $n_{avg}$. Fluorescence light collection is typically low due to the diamond's high refractive index causing light trapping. Just recently up to 95% collection efficiency has been demonstrated [56]. Engineering endeavor, but recommended for single-cell NV-NMR. | ~3 fold compared to [16,18] | Unaffected |
| | | Quantum logic enhanced readout | Improves $n_{avg}$. Information of the NV spin state is destroyed during the readout process due to concomitant repolarization into the $|0\rangle$ state. QLE schemes map the NV spin state to an ancillary nuclear spin for repetitive readout. Commonly employed for single NV centers and recently demonstrated for NV ensembles [41]. Prospects for single-cell NV-NMR considered very promising. | ~10 fold compared to [16,18] | Unaffected |
| | Diamond material | Optimized NV density and NV properties | Improves C, $T_2$ and $N_{NV}$. Increasing the N-to-NV conversion efficiency at constant nitrogen concentration [N] can reduce charge state stability and contrast C [42]. Assuming constant N-to-NV conversion efficiency, increasing the NV density has little effect on the sensor's sensitivity for $[N] \geq 0.8\ ppm$, since $[N] \propto 1/T_2$ [57]. Nevertheless, due to technical difficulties arising from high NV densities and favourable scaling of QLE readout schemes with long NV coherence times, low [N] are typically preferred. Co-doping of non-nitrogen-based n-type dopants, such as sulfur or phosphorous [44–46], has led to improvements in NV center conversion efficiencies, coherence time [45] and charge state stability [44]. However, more research is needed to explore the full potential. Prospects for single-cell NV-NMR considered promising. | ~5 fold compared to [16,18]; more research needed | Unaffected |
| NMR signal | NMR linewidth | Enhanced magnetic field uniformity and advanced NMR pulse sequences | The magnetic field uniformity can be increased through the use of homogeneous magnets and shimming [49]. In addition, advanced NMR pulse sequences can be used to narrow linewidths [58]. We expect that state-of-the-art benchtop NMR standards can be reached, corresponding to $\leq 0.5$ Hz linewidth at 1 T. Mainly an engineering endeavor, prospects for single-cell NV-NMR considered very promising. | sample dependent; $T_1$ relaxation time determines the experiment repetition rate | Line-broadening due to magnetic susceptibility mismatches possible, in particular at high magnetic fields |
| | Nuclear spin polarization | Increased magnetic field strength | Nuclear spin polarization is directly proportional to the magnetic field strength and, therefore, the NMR signal strength increases linearly with the magnetic field. Challenging due to increasing complexity of MW equipment needed for the NV-spin manipulation with increasing magnetic field. Reaching 1 T is realistic. Prospects for single-cell NV-NMR considered promising. | ~11 fold (1 T) compared to [16] | Field dependent line-broadening due to magnetic susceptibility mismatches |
| | | PHIP | Low complexity hyperpolarization technique that is fast, cheap and offers high polarization levels (25% $^{13}$C polarization after purification have been demonstrated [53]), however, the method currently suffers from a low generality and toxic contaminants. Prospects for single-cell NV-NMR are very promising. | ~12 fold ($^{13}$C, lower gyromagnetic ratio considered for calculation) compared to [18] | Cells have to take up and metabolize hyperpolarized species; experiments limited by $T_1$ |
| | | dDNP | A general hyperpolarization technique reaching extremely high polarization levels (70% $^{13}$C liquid-state polarization [55]), however, its widespread use is limited by high costs, high complexity, contaminants and very low throughput/long polarization times. Prospects for single-cell NV-NMR are very promising. | ~35 fold ($^{13}$C, lower gyromagnetic ratio considered for calculation) compared to [18] | Cells have to take up and metabolize hyperpolarized species; experiments limited by $T_1$ |

**Microscale NV-NMR on biological samples**

While improvements of the microscale NV sensor's sensitivity are generally unaffected by biological samples, improvements arising from NMR parameters can be negatively impacted. Line-broadening due to magnetic susceptibility mismatches can reduce the achieved LOD, especially in the untargeted approach, where high magnetic fields are beneficial. The targeted approach is mainly limited by the relaxation times $T_1$ of the hyperpolarized metabolic probes, which have to be taken up by cells and rapidly metabolized to efficiently study metabolic fluxes.

For studying biological samples the biocompatiblity of NMR spectroscopy is a vital aspect and a prerequisite. In contrast to inductive NMR techniques NV-NMR requires high laser powers and strong MW fields. Crucially, a microfluidic platform for NV-NMR has been realized, utilizing a total internal reflection geometry, which limits laser-induced photodegradation [36]. This platform could be adapted for single-cell NV-NMR incorporating cell culturs or cell traps. Given the significant amount of water in biological samples, strong MW fields may cause sample heating, ultimately leading to cell death. This effect could be reduced by the use of MW resonators, decreasing the electric field component within the sample volume.

**Conclusion**

In recent years, NMR spectroscopy at the microscale has made considerable progress due to rapid developments in the field of NV-based quantum sensing. These atom-sized sensors were used to probe picoliter volumes, i.e., volumes of small eukaryotic cells. Despite this significant progress, the concentration sensitivity and chemical specificity for single-cell analysis must still be improved. Therefore, we provided a possible roadmap towards single-cell NV-NMR by combining recent advances in the fields of NV-based quantum sensing and NMR spectroscopy to predict the achievable sensitivities. We envision that in an untargeted approach the detection of Boltzmann nuclear polarization will enable studying of intracellular water diffusion [37] or single-cell spectroscopic fingerprinting of highly abundant metabolites. In combination with machine learning approaches, it may allow to classify a cell's phenotype. Using a targeted approach in which isotopically labeled and hyperpolarized metabolites are fed to cells and detected, we expect microscale NV-NMR to achieve attomol nLOD at picoliter volumes corresponding to micromolar cLOD and to become a powerful tool for fluxomic studies of small eukaryotic and even prokaryotic cells. Monitoring the actual reaction rates in the metabolic networks of individual cells will enable the identification of metabolic pathways with applications in biotechnology or pharmacology ranging from metabolic modeling to screening of metabolic variants [59–61].

**Author Contributions**

N.R.N. and D.B.B. conceptualized and wrote the manuscript. R.D.A. revised the manuscript.

**Acknowledgements**

This work was supported by the European Research Council (ERC) under the European Union's Horizon 2020 research and innovation program (grant agreement No 948049) and by the Deutsche Forschungsgemeinschaft (DFG; German Research Foundation) under Germany's Excellence Strategy EXC-2111 390814868.

**Declaration of Competing Interests**

N.R.N, R.D.A. and D.B.B. declare that they have no known competing financial interests or personal relationships that could have appeared to influence the work reported in this article.

**Reference Annotations**

**\*\* Glenn Nature 2018**

In this article, NV-NMR at the microscale using a several micrometer thick layer of NV ensembles is demonstrated for the first time. Furthermore, a novel quantum sensing scheme is presented, that allows to probe NMR signals from picoliter volumes with high frequency resolution and therefore overcomes the spectral resolution limit imposed on commonly used NV sensing schemes caused by NV decoherence and short spin state lifetimes.

**\*\*Allert Lab on a chip 2022**

In this article, a fully integrated microfluidic platform for NV quantum sensors is presented. The microfluidic platform is highly adaptable and biocompatible.

**\*Smits Sci. Adv. 2019**

In this article, two-dimensional NV-NMR spectroscopy of liquid analytes in an effective sensing volume of 40 picoliter is demonstrated.

**\*Arunkumar PRX Quantum 2021**

In this article, microscale NV-NMR is combined with SABRE hyperpolarization achieving a record proton number sensitivity of $66 \, fmol/\sqrt{Hz}$.

**\*\* Barry Rev. Mod. Phys. 2020**

This review provides an excellent in-depth analysis of strategies to improve the NV magnetic field sensitivity.

**\*\*Degen Rev. Mod. Phys. 2017**

This review provides an excellent introduction to the basic principles, concepts and methods of quantum sensing.

**\*Bruckmaier JMRO 2021**

In this article, the dependence of the NMR signal size on the diamond sensor's and sample's geometry is discussed in detail.

**\*\*Allert ChemComm 2022**

This review provides a good introduction to the basic principles of nanoscale and microscale NV NMR spectroscopy.

*Bruckmaier Arxiv 2023

In this article, microscale NV NMR is combined with the pulsed gradient spin echo sequence to study molecular diffusion and flow in microstructures.